\documentclass[conference,10pt]{IEEEtran}
\IEEEoverridecommandlockouts
\usepackage[normalem]{ulem}

\usepackage[hyphens]{url}
\usepackage[hidelinks, breaklinks]{hyperref}
\usepackage{cite}
\usepackage{graphicx}
\usepackage{booktabs}
\usepackage{csquotes}
\usepackage{tikz}

\usetikzlibrary{arrows.meta, positioning, fit, backgrounds, shapes.geometric,
                shapes.symbols, shapes.arrows, shadows, calc}
\usepackage{fontawesome5}
\usepackage[capitalize]{cleveref}
\usepackage{xspace}
\usepackage{subcaption}
\usepackage{enumitem}

\def\eg{\textit{e.g.},\xspace}

\begin{document}

%\title{Position Paper: Towards an Internet of Collaborating Things and the Role of Agentic Edge AI}
\title{
The Internet of Collaborating Things: Agentic Edge AI for Autonomous Cross-Domain Collaboration}

%Toward Agentic Orchestration for the Internet of Collaborating Things}

%% Double-blind: author block anonymized for review
% \author{\IEEEauthorblockN{Position Paper (6 Pages)}}
\author{%
\IEEEauthorblockN{%
Walid A. Hanafy\IEEEauthorrefmark{1}, %
Nader Sehatbakhsh\IEEEauthorrefmark{2}, %
David Irwin\IEEEauthorrefmark{1}, %
Mani Srivastava\IEEEauthorrefmark{2}, % 
Prashant Shenoy\IEEEauthorrefmark{1}%
}

\vspace{.15cm}
\IEEEauthorblockA{\parbox{\linewidth}{\centering
\IEEEauthorrefmark{1}University of Massachusetts Amherst, 
\IEEEauthorrefmark{2}University of California, Los Angeles
}}
}
\maketitle

\begin{abstract}
The Internet of Things is on a trajectory toward a trillion connected devices
deployed across multiple domains. These devices are no longer simple sensing
and actuation endpoints; they are mobile platforms with embedded processing
and intelligent on-device services. 
The dominant paradigm of offloading
computation to cloud and edge servers is a vertical device-to-server
interaction model that cannot scale with this trajectory.
What is needed
instead is a horizontal paradigm in which devices communicate and collaborate
directly, pooling their compute, sensing, and actuation into dynamic,
cross-domain clusters rather than offloading to centralized infrastructure.
We refer to this paradigm as the Internet of Collaborating Things (IoCT). 
Realizing this vision requires both a portable and secure execution substrate for
heterogeneous hardware and an agentic control plane capable of contextual
reasoning, transient trust establishment, and open-world adaptation throughout
the device collaboration lifecycle without human intervention. 
In this position paper, we define this new communication, compute, and collaboration paradigm,
articulate the role of agentic edge AI in realizing it, and outline research
challenges and future directions toward trustworthy, autonomous IoCT
collaboration.

\end{abstract}


\section{Introduction}
\label{sec:introduction}

\begin{figure}
  \begin{subfigure}[b]{0.7\linewidth}
   \includegraphics[width=\linewidth]{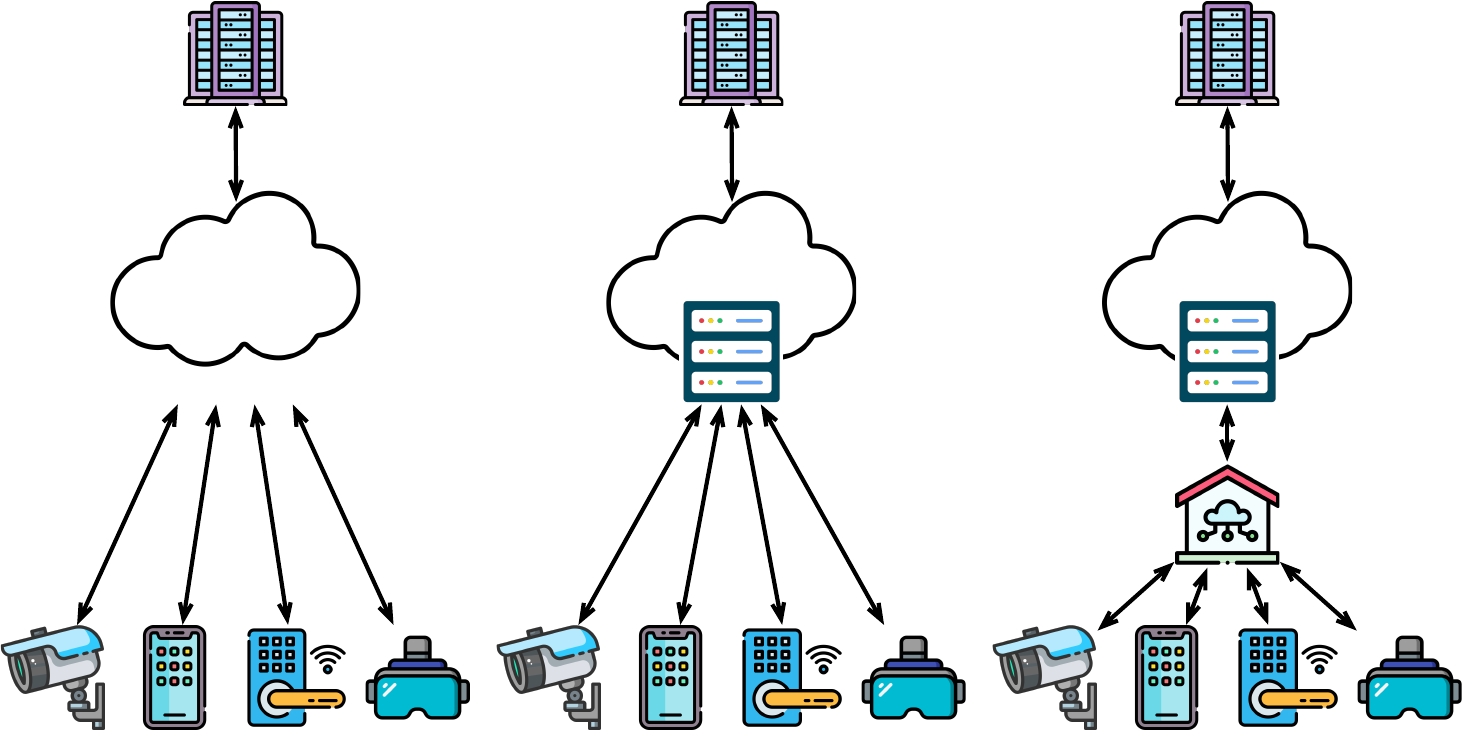}
   \caption{On-premise-edge-cloud hierarchy}
   \label{fig:paradigm_vertical}
  \end{subfigure}
   \hfil
   \begin{subfigure}[b]{0.255\linewidth}
   \includegraphics[width=\linewidth]{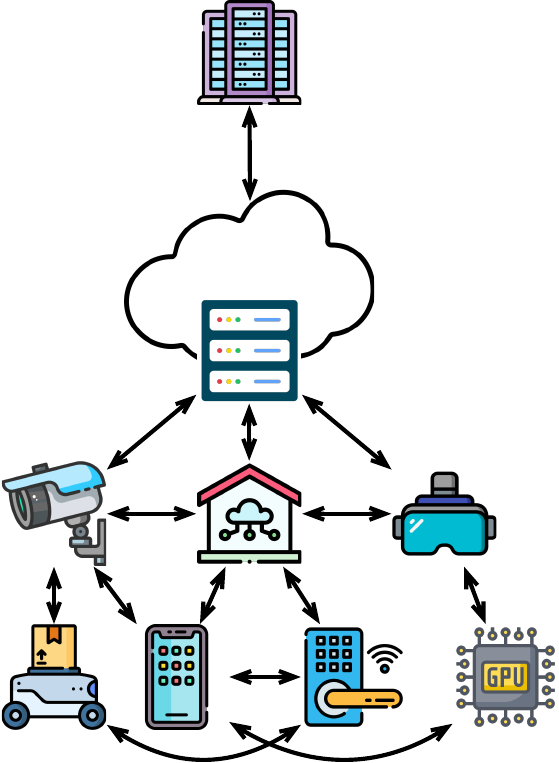}
   \caption{IoCT}
  \label{fig:paradigm_IoCT}
  \end{subfigure}
  \vspace{15pt}
  \caption{From a \textbf{vertical} offload paradigm, in which devices
  funnel data up through an on-premise-edge-cloud hierarchy (a), to a
  \textbf{horizontal} \emph{Internet of Collaborating Things}, in which
  heterogeneous devices communicate and collaborate directly, pooling their compute, sensing, and actuation capabilities (b).}
  \label{fig:paradigm}
\end{figure}

The number of connected Internet of Things (IoT) devices exceeded
20~billion in 2025~\cite{IoTAnalytics2025} and some envision an Internet of a Trillion Things (IoTT) within the next decade~\cite{IoTTrillion}.
Moreover, with recent advances in embedded hardware, low-power architectures, and domain-specific acceleration, IoT devices are becoming increasingly capable~\cite{Xu2014:IoT_Industry}.
Modern IoT devices integrate multicore processors, AI accelerators, cameras, and LiDAR, often matching or exceeding the compute
capabilities of the on-premise gateways that are supposed to serve
them~\cite{Gupta2019:Chiplets,Shi2016:EdgeComputingVision,Satya2017:Edge,Lin2020:MCUNet}.
Despite this increase in device capabilities, today's IoT infrastructure is still organized around \emph{offloading}, a device-to-server interaction model in which devices generate data and
delegate processing to a tiered hierarchy of on-premise, edge, and cloud
servers~\cite{Shi2016:EdgeComputingVision,Satya2017:Edge,aws-iot,azure_iot_edge_2026}
(see \cref{fig:paradigm_vertical}). With continued growth, the number of IoT devices will dwarf the number of servers available to process their data by many orders of magnitude. Consequently,
the current offloading-based \emph{vertical} interaction model will not be able to sustain the growth in the number of IoT devices as well as the processing and bandwidth requirements of modern AI applications and high-resolution sensing modalities~\cite{Shi2016:EdgeComputingVision, Meng2022:DoWeNeedEdge}. 
This pressure is further amplified by the emergence of Physical AI workloads, including embodied agents, collaborative robotics, multimodal perception systems, and AR/VR platforms. These workloads increasingly require low-latency coordination across distributed sensors, actuators, and accelerators in the physical environment, while generating data volumes and inference demands that cannot be sustainably funneled through centralized infrastructure alone.

These issues are exacerbated for mobile and transient devices, \eg drones and robots navigating unfamiliar territories, frequent visitors, and collaborative tasks, as detailed in \cref{sec:ioct_examples}. In these scenarios, because devices interact with nearby devices and leverage their compute, sensing, and actuation capabilities, a centralized cloud or edge \emph{vertical} interaction approach will incur an unnecessarily high performance overhead. Instead, this position paper argues for a new paradigm that enables \emph{horizontal} interactions between nearby devices, pooling the compute, sensing, and actuation capabilities they
already carry (\cref{fig:paradigm}b).

We refer to this new paradigm as the \emph{Internet of Collaborating Things}
(IoCT). IoCT embraces the dispersed availability and sensing capabilities
of IoT devices, and builds \emph{dynamic}, \emph{cross-domain}
collaborative clusters that extend the conventional cloud-edge continuum
downward to the device layer.
%\walid{I don't think we need to replace this model; we should augment or extend it.}
In this model, the IoT and gateway devices in a physical space constitute a shared
resource pool, harnessing the sensing, actuation, computing, and
networking capabilities of co-located devices. This pool is available to
multiple applications and visiting devices simultaneously.
Importantly, cross-device collaboration and the associated
\emph{perception}--\emph{cognition}--\emph{action} loops get organized
(and reorganized) across an amorphous collection of IoT devices that span
different trust domains.  

In IoCT, each application sees a dynamic and heterogeneous cluster of IoT
devices and on-premise servers, with wide-area connectivity to edge and
cloud infrastructure. To enable seamless cross-device collaboration and
cooperative execution, IoCT applications are composed as portable dataflow
graphs whose leaf nodes are either \emph{nanotasks}, platform-independent
computation mapped at runtime to the CPUs, GPUs, and accelerators available
on nearby devices~\cite{Kotsifakou2018:HPVM}, or \emph{nanoservices} that use sensing, actuation, and communication primitives exposed by the different devices.

Providing this flexible, on-demand computational substrate raises four
interrelated challenges:
\begin{enumerate}[leftmargin=!]
\item \textbf{Device Discovery.} The IoCT environment is diverse and fluid. Thus, the substrate must
continuously accommodate devices as they arrive and depart, identify their capabilities,  
maintain an up-to-date network view, and auto-configure devices and networks for seamless access, all without human intervention.
\item \textbf{Cross-Domain Security and Access Control.} Because devices roam across administrative domains, every collaboration session requires authenticating visiting devices, dynamically attesting \textit{both} code and shared data, and enforcing fine-grained access control policies that are generated on demand, scoped to a specific interaction, and revoked upon departure.

\item \textbf{Heterogeneity-Aware Orchestration.} 
Unlike homogeneous edge servers, the device layer incorporates a
high degree of hardware heterogeneity, making it challenging to determine
how nanotasks should be mapped to devices. Further, as workloads and the device population change rapidly, the substrate must dynamically adapt how
resources are allocated to nanotasks and nanoservices.

\item \textbf{Constrained Execution.}
Beyond mapping nanotasks to heterogeneous devices, the substrate must
execute compute-intensive and latency-sensitive workloads under tight
compute, memory, energy, and thermal constraints. This challenge is
especially pronounced for AI workloads, where models and
agents may need to be partitioned, compressed, migrated, or adapted at
runtime to balance accuracy, latency, energy consumption, and availability.

\end{enumerate}

Realizing IoCT and overcoming these challenges will require rethinking current edge architectures, including their data and control planes.
On the \emph{data
plane}, IoCT applications require a portable execution substrate that maps nanotasks to heterogeneous
hardware~\cite{Haas2017:WebAssembly,Kotsifakou2018:HPVM,Noor2019:DDFlow} and provides an adaptive, trusted, and secure execution platform~\cite{wang2022rt, alder2021aion, arkannezhad2024ida, Sehatbakhsh2019:EMMA}, as well as ensuring secure and trusted communication and access control between IoCT devices from different administrative domains~\cite{Shakarami2022ScenarioDrivenDA, Liu2021:Aerogel, Alkhresheh2020:DACIoT}.
On the
\emph{control plane}, a unified autonomous orchestration layer must discover
devices, validate cross-domain credentials and attestations, allocate
resources, and adapt continuously as clusters form and dissolve in open-world
physical environments~\cite{Shastri2025:LLM-Driven-CollabIoT, zeroconf, ccori2016device, achir2022service}.

\noindent\textbf{Our Position.} Realizing the IoCT vision requires a fundamentally new edge
architecture, one governed by autonomous agents that manage the full device
collaboration lifecycle across administrative trust domains without human
intervention at runtime. These agents provide the contextual reasoning,
transient trust establishment, and open-world adaptation needed when devices
arrive with unfamiliar identities, capabilities, policies, and task needs. In
this architecture, human administrators express high-level intent, while agents
inspect newly discovered devices' identities, capabilities, and task needs, and
evaluate credential and attestation results.
Agents then produce validated policy and placement decisions and adapt resource
allocations as devices arrive, depart, or move through a space. This allows devices to shift from standalone operation to
cooperative use of nearby resources when context and policy permit, without
requiring manual configuration for each encounter. Without this autonomy,
IoT orchestration cannot scale to the frequency, heterogeneity, and
cross-domain mobility expected in future trillion-device deployments.

This position paper makes three contributions. First, we define the IoCT paradigm, present motivating scenarios, and show why existing approaches fall short (\cref{sec:ioct}). 
Second, we propose an initial IoCT architecture with a data plane for portable cross-device execution and an agentic control plane for autonomous orchestration (\cref{sec:realizing}). 
Third, we identify open research challenges spanning verifiable agentic decisions, trust negotiation, and resource management (\cref{sec:challenges}).

\begin{figure*}[t]
   \hfil
  \begin{subfigure}[b]{0.3\linewidth}
   \includegraphics[width=\linewidth]{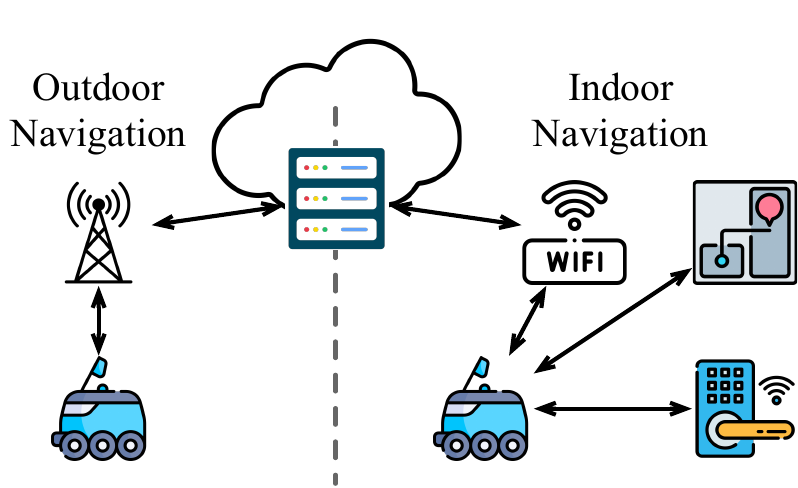}
   \caption{Delivery Robot}
   \label{fig:examples_robot}
  \end{subfigure}
   \hfil
   \begin{subfigure}[b]{0.25\linewidth}
   \includegraphics[width=\linewidth]{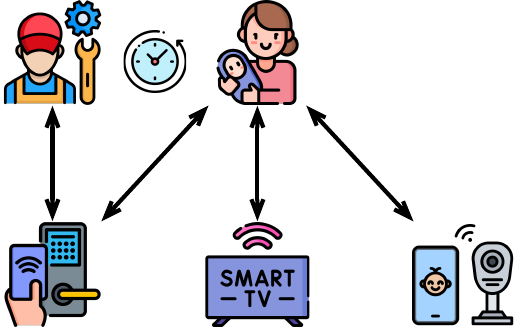}
   \caption{Frequent Visitors}
   \label{fig:examples_visitor}
  \end{subfigure}
   \hfil
  \begin{subfigure}[b]{0.35\linewidth}
   \includegraphics[width=\linewidth]{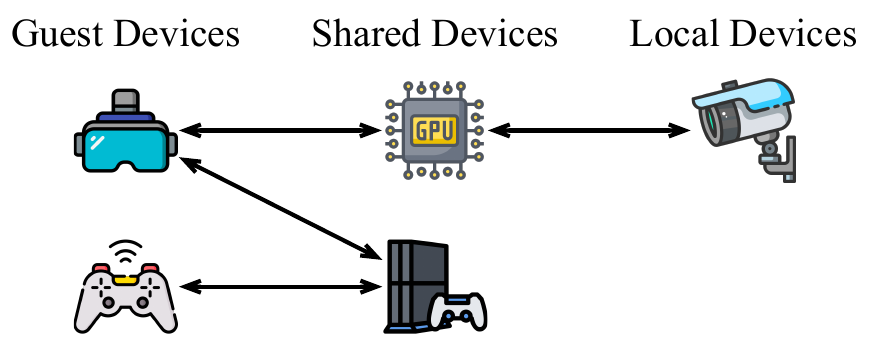}
   \caption{Collaborative Gaming}
   \label{fig:examples_gaming}
  \end{subfigure}
  \caption{Three motivational scenarios for IoCT. In each case, visiting
  devices enter a foreign space and collaborate with resident devices
  across trust domain boundaries.}
  \label{fig:examples}
\end{figure*}

\section{Why Is IoCT Hard to Achieve?}
\label{sec:ioct}

This section presents motivating examples for IoCT and explains why the state of the art falls short.

\subsection{Motivational Examples}\label{sec:ioct_examples}
We begin with future IoT scenarios where the conventional vertical offloading model shown in \cref{fig:paradigm_vertical} is no longer adequate and IoCT offers a more effective alternative.
The following scenarios illustrate what this horizontal paradigm looks like in practice.

\smallskip\noindent\textbf{Delivery Robot.}
Food delivery robots (see~\cref{fig:examples_robot}) are now common on many university campuses~\cite{ucla_robot_delivery}, and delivery providers and online retailers are increasingly experimenting with autonomous robots or drones to deliver packages~\cite{doordash2025dot, amazon_prime_air}.
A robot may navigate \emph{autonomously} outdoors using local vision, LiDAR, GPS, and optional computation offload to 5G-integrated
edge servers such as AWS Wavelength~\cite{aws_wavelength_2026}. Once near or inside a building, the robot can switch from \emph{standalone} navigation to \emph{cooperative} operation with the building's available sensors and devices.
To do so, the robot must connect to the building's Wi-Fi network, query its indoor map, request authorized entry through smart-lock services~\cite{amazon_key}, and place computation on on-premise servers to benefit from lower latency,
improved reliability, and higher-quality prediction and control.
This scenario poses challenges for secure multitenancy, trustworthy resource pooling across trust domains, data integrity, and performance and latency management that are not addressed by current offloading-focused paradigms.

\smallskip\noindent\textbf{Frequent Visitors.}
Another example is a frequent visitor (see~\cref{fig:examples_visitor}), such as a babysitter or maintenance personnel, who requires access to the environment's devices.
The visitor's device must be granted access to the door locks and internal appliances relevant to their expected duties.
For example, the babysitter requires access to the baby monitor camera and limited access to entertainment devices (\eg a TV or wireless speakers).
This scenario poses challenges for fine-grained, time-bounded access control that are not addressed by current offloading-focused paradigms~\cite{He2018:Rethinking}. For privacy reasons, the babysitter should be able to access the baby monitor only during the visit, without being able to reconfigure it.

\smallskip\noindent\textbf{Collaborative Gaming.} Consider a group of friends who meet at one of their homes to play together (see~\cref{fig:examples_gaming}). They bring their AR headsets (\eg Apple Vision Pro~\cite{Apple_vison_pro}), wireless game controllers, and mobile devices, and gain access to the home's smart TV and gaming console for multiplayer gaming, while controlling the living room's lights and music for an enhanced experience. The AR headsets can also use an on-premise PC or server for AR processing. This scenario requires applications to coordinate resources and services while maintaining performance isolation among users, applications, and background home services.

\smallskip
These scenarios span different physical domains, device types, and
application requirements, yet they share a common structure: a
\emph{visiting} device enters a \emph{space} (an administrative domain
such as a building or hospital ward) that governs a population of
\emph{resident} devices. The visiting device discovers available resources,
obtains scoped access across \emph{trust domain} boundaries,
collaborates, and eventually departs. We use the delivery robot as an example throughout this position paper.

\subsection{Limitations of State of the Art}\label{sec:limitations}
Current approaches fall short along one or more IoCT requirements.

\smallskip\noindent\textbf{Cloud and Edge Platforms.}
Commercial IoT platforms such as AWS IoT~\cite{aws-iot}, Azure IoT
Edge~\cite{azure_iot_edge_2026}, and AWS
Greengrass~\cite{AWSIoTGreengrass} position
computation across the cloud--edge--server continuum. AWS
Wavelength~\cite{aws_wavelength_2026} pushes services onto 5G edge nodes for
lower latency. 
However, these platforms treat devices as clients
of infrastructure, not as peer collaborators. 
Device registries are
pre-provisioned, trust is anchored in a single vendor's identity
system, and there is no mechanism for cross-domain resource pooling.

\smallskip\noindent\textbf{IoT Composition Frameworks.}
Systems for composing IoT applications across distributed
devices, such as declarative dataflow frameworks~\cite{Noor2019:DDFlow}
and service-oriented IoT
middleware~\cite{Giang2015:Distributed}, partition applications
statically at deployment time. They do not handle dynamic device
populations, visiting devices from foreign domains, or runtime
reallocation as the environment changes.

\smallskip\noindent\textbf{Secure Isolation Mechanisms.}
WebAssembly (WASM) provides portable, sandboxed execution and has been
shown viable on microcontrollers~\cite{Haas2017:WebAssembly}, but its runtime lacks
support for real-time scheduling, multi-tenant sensor sharing, and
hardware accelerator access. Trusted execution environments (TEEs) such
as ARM TrustZone and Intel SGX offer hardware-backed isolation but are
hardware-specific, unavailable on many low-end devices, and do not
address cross-domain authentication, attestation validation, or scoped
access-control decisions by themselves~\cite{Brasser2015:TyTAN}.

\smallskip\noindent\textbf{Attestation and Device Integrity.}
Remote attestation methods, whether hardware-based (TPMs, SGX enclaves),
software-based, or hybrid~\cite{Brasser2015:TyTAN,noorman2017sancus}, can verify
device integrity, but are designed for static deployments with
pre-established verifier--prover relationships. They do not support the
transient, cross-domain authentication and attestation workflows that feed
scoped access-control decisions. Privacy-preserving
techniques such as differential privacy~\cite{Dwork2014:AlgorithmicPrivacy} and
homomorphic encryption~\cite{Acar2018:HESurvey} can limit data disclosure in
specific processing settings, but do not establish dynamic data or code integrity or
guarantee the trustworthiness of sensor inputs or actuator outputs. 

\smallskip\noindent\textbf{Service Discovery.}
Protocols such as UPnP~\cite{ocf_upnp_arch_2020}, Zero-Conf~\cite{zeroconf}, and
Jini~\cite{Waldo2000:Jini} provide plug-and-play device registration within a
local network but assume implicit trust among participants. They offer no
access control, capability profiling, or cross-domain identity
verification, and are not designed for untrusted or multi-tenant
environments.

\section{Realizing IoCT}
\label{sec:realizing}

Before presenting the architecture and the role of agentic edge AI, we list the core design principles guiding IoCT:

\begin{itemize}[leftmargin=*]
    \item \textbf{Horizontal collaboration over hierarchical offload.}
    Rather than treating IoT devices solely as clients of edge or cloud infrastructure, IoCT enables nearby devices to collaborate directly and pool sensing, computation, networking, and actuation resources.

    \item \textbf{Transient cross-domain federation.}
    IoCT collaborations are dynamic and short-lived, forming opportunistically as devices enter a physical space and dissolving when they depart, often across administrative trust boundaries with no prior relationship.

    \item \textbf{Portable execution across heterogeneous devices.}
    Applications must execute seamlessly across diverse hardware platforms, including microcontrollers, embedded processors, GPUs, and specialized accelerators, without requiring device-specific deployment logic.

    \item \textbf{Policy-driven autonomous orchestration.}
    Because IoCT environments are too dynamic and heterogeneous for manual management, orchestration must be driven by high-level intent and automated reasoning over device capabilities, trust relationships, and runtime context.

    \item \textbf{Continuous adaptation to mobility and resource dynamics.}
    The system must continuously reconfigure placement, access control, and resource allocation as devices move, workloads fluctuate, and network conditions evolve.

    \item \textbf{Least-privilege, time-bounded collaboration.}
    Access to resources and services should be scoped to the minimum permissions required for a specific collaboration session and revoked automatically when the interaction terminates.
\end{itemize}
%}

\subsection{An Initial IoCT Architecture}
\label{sec:ioct-arch}

The IoCT architecture considers heterogeneous IoT devices in a physical
space, including resident and visiting devices
that may belong to different administrative domains. These devices vary in
compute, sensing, and actuation capabilities, and some have wide-area
connectivity to edge and cloud infrastructure.
IoCT uses two complementary layers.
A portable and secure communication and execution substrate handles
heterogeneous hardware, while a new control plane orchestrates the full
device collaboration lifecycle without human intervention.

\smallskip\noindent\textbf{Data Plane.}
Rather than treating devices only as clients of the edge-cloud continuum,
an application sees a dynamic and heterogeneous cluster of IoT devices and
on-premise servers with wide-area network connectivity to edge and cloud
servers. 
IoT devices have sensors,
actuators, embedded processors, and specialized accelerators, while on-premise, edge, and cloud servers
offer general-purpose computing resources, including CPUs, GPUs, and FPGAs. 
The IoCT architecture
enables IoT devices to discover and leverage the sensing, actuation, computation, and connectivity available from devices and servers within the visited space. 

Device collaboration requires a communication substrate that spans the heterogeneous wireless links (Wi-Fi, 5G/6G, BLE~\cite{schussmann2017bluetooth}, LoRa~\cite{Devalal2018:LoRaComm}) present in a typical physical space. 
Rather than requiring applications to manage each link individually, the data plane should provide a unified messaging abstraction, such as a publish-subscribe model, that bridges communication across device boundaries transparently.
For example, an embedded message broker on each device can provide local module-to-module messaging, while brokers on different devices bridge to one another over the network using lightweight IoT protocols such as MQTT~\cite{MQTT5}. 
The broker also mediates access control by checking capability-based tokens at wire speed~\cite{Shastri2025:LLM-Driven-CollabIoT}, ensuring that policies generated by the control plane are enforced on every interaction. 

In addition to communication, the data plane supports a hardware-independent deployment substrate. 
For portable execution, IoCT can use
WebAssembly (WASM), which provides a
language- and hardware-independent binary format that offers compact
representation, efficient validation, and safe low-overhead
execution~\cite{Haas2017:WebAssembly}, and has been shown to be viable on
resource-constrained microcontrollers. To protect not only the host
device from misbehaving tenant code, but also the tenant's code from a
misbehaving host, the WASM runtime can be placed inside a hardware
trusted execution environment (TEE) such as ARM
TrustZone~\cite{menetrey2022watz,yuhala2024fortress}, creating a two-way isolation
boundary. Beyond CPU and memory, the data plane must also support
fine-grained, multi-tenant sharing of sensors, actuators, and hardware
accelerators, despite the lack of conventional virtualization for these IoT devices.

\begin{figure}[t]
  \centering
  \includegraphics[width=0.8\linewidth]{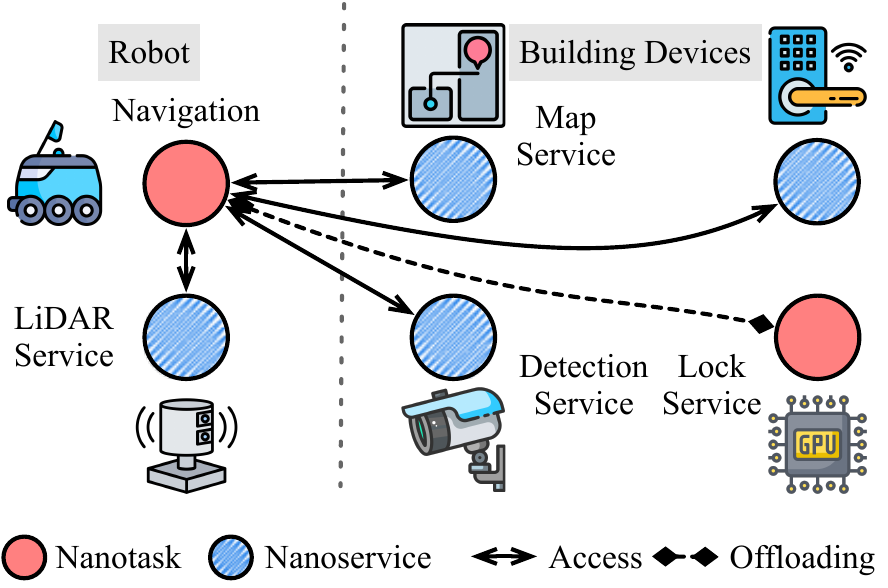}
  \caption{Dataflow graph for the delivery-robot scenario. The robot's
  mission decomposes into nanotasks and nanoservices mapped across two trust domains.}
  \label{fig:robot_flow}
\end{figure}

\smallskip\noindent\textbf{Application Model.}
An IoCT application is expressed as a portable dataflow graph whose
nodes are either \emph{nanotasks} or \emph{nanoservices}. 
%\sout{Nanotasks are platform-independent units of computation that can be mapped at runtime to the CPUs, GPUs, or accelerators available on nearby devices~\cite{Kotsifakou2018:HPVM}, while nanoservices are sensing, actuation, and communication primitives exposed by resident and visiting devices alike.}
Nanotasks are lightweight, platform-independent computation units that can be dynamically placed across nearby heterogeneous devices at fine granularity~\cite{Kotsifakou2018:HPVM}. Nanoservices expose device-native sensing, actuation, and communication primitives as composable runtime services.

To enable this mapping, each device advertises its available resources
and nanoservices using structured capability descriptions (\eg
protocol-buffer payloads over MQTT), allowing the runtime to discover
available hardware and services.

In the delivery-robot scenario, the robot can use its own
resources as well as resources provided by the building. 
As shown in \cref{fig:robot_flow}, the robot's mission can be
fulfilled using a set of nanotasks (\eg navigation), onboard services, and
building-provided services. For instance, the navigation task
can use onboard LiDAR to plan a path, and use
building nanoservices, including a map nanoservice that returns the
building layout or a lock-actuation nanoservice that triggers the
apartment's smart lock upon successful verification. The robot can
execute these tasks locally or use the building's
resources, based on device capabilities, resource availability, and access
permissions.

\smallskip\noindent\textbf{Control Plane.}
Complementary to the data plane, a control plane is needed to decide
\emph{which} device runs \emph{what}, under \emph{whose} authority, and
to \emph{adapt} those decisions as conditions change. This requires
several interrelated functions: \emph{(i) discovery and capability
profiling:} detecting new devices and learning what resources and
services they offer, extending zero-configuration
networking to cross-domain
settings~\cite{zeroconf}, \emph{(ii) resource management:} mapping nanotasks onto the
available device pool under latency and capacity constraints, and
re-mapping when devices arrive or depart, \emph{(iii) cross-domain
authentication and attestation:} authenticating visiting devices and
validating credentials and attestations across administrative
domains, \emph{(iv) fine-grained access
control:} generating and enforcing time-bounded policies that govern
which devices may use which resources, and \emph{(v) runtime
adaptation:} continuously monitoring and adjusting all of the above as the physical context evolves.

To realize these functions, the control plane maintains a device registry that tracks the capabilities, locations, and credential and attestation metadata of all devices in the space, enabling both discovery and resource allocation. 
When a visiting device from a different trust domain arrives, the control plane performs credential exchange and attestation checks for that visit.
The validation results inform the fine-grained access-control policies for the visit, which are encoded as short-lived capability tokens and pushed to the data-plane message brokers for enforcement at wire speed. 
The control plane maps nanotasks onto available devices under latency and capacity constraints, and continuous monitoring triggers re-orchestration when the context changes, whether a device departs, a workload shifts, or a visiting device moves to a different zone. 
This separation keeps reasoning and control decisions in the control plane
while leaving low-latency enforcement to the data plane.
However, because the device combinations and contexts involved cannot be
anticipated at design time, and new collaboration sessions arise too
frequently for manual configuration, the control plane must operate
autonomously. We propose an agentic control plane in which agents
reason over intent, context, trust evidence, and resource state to handle these
functions.

\subsection{Leveraging Edge AI Agents}
\label{sec:architecture}

The IoCT control plane must continuously discover devices, validate
cross-domain credentials and attestations, generate policies, and adapt to
changing conditions, all without
human intervention at runtime. Static rules or pre-programmed logic cannot address these tasks because the device combinations, credential
requirements, and environmental conditions are impossible to anticipate at
design time. 
Further, the challenge is not merely optimization complexity, but open-world contextual reasoning under incomplete and continuously evolving information, where rigid rule-based orchestration becomes brittle and unscalable.
Thus, the IoCT control plane requires an \emph{agentic orchestrator} that can
perform contextual reasoning, establish transient trust, and adapt in open-world
physical environments. Recent LLM-based agent architectures provide one
promising way to implement this reasoning layer~\cite{Wei2022:CoT,yao2023:React,Sumers2024:CoALA, Shen2023:HuggingGPT, Wang2024:LLMAgentSurvey}.
The agent receives
high-level intent from human administrators, and supports the collaboration
lifecycle by producing validated access-control, placement, and
reconfiguration decisions for the control plane to apply. This builds on the
LLM-as-orchestrator paradigm~\cite{Shen2023:HuggingGPT}, but extends it with
continuous reactivity and domain-grounded validation.
The orchestrator maintains a persistent observe-reason-act loop over the
physical environment rather than executing a one-shot plan.
The IoCT orchestrator is event-driven, remaining idle until the control plane raises an alert and then reasons over the current state to produce a targeted
decision rather than re-planning from scratch.

\begin{figure}[t]
  \centering
  \includegraphics[width=\linewidth]{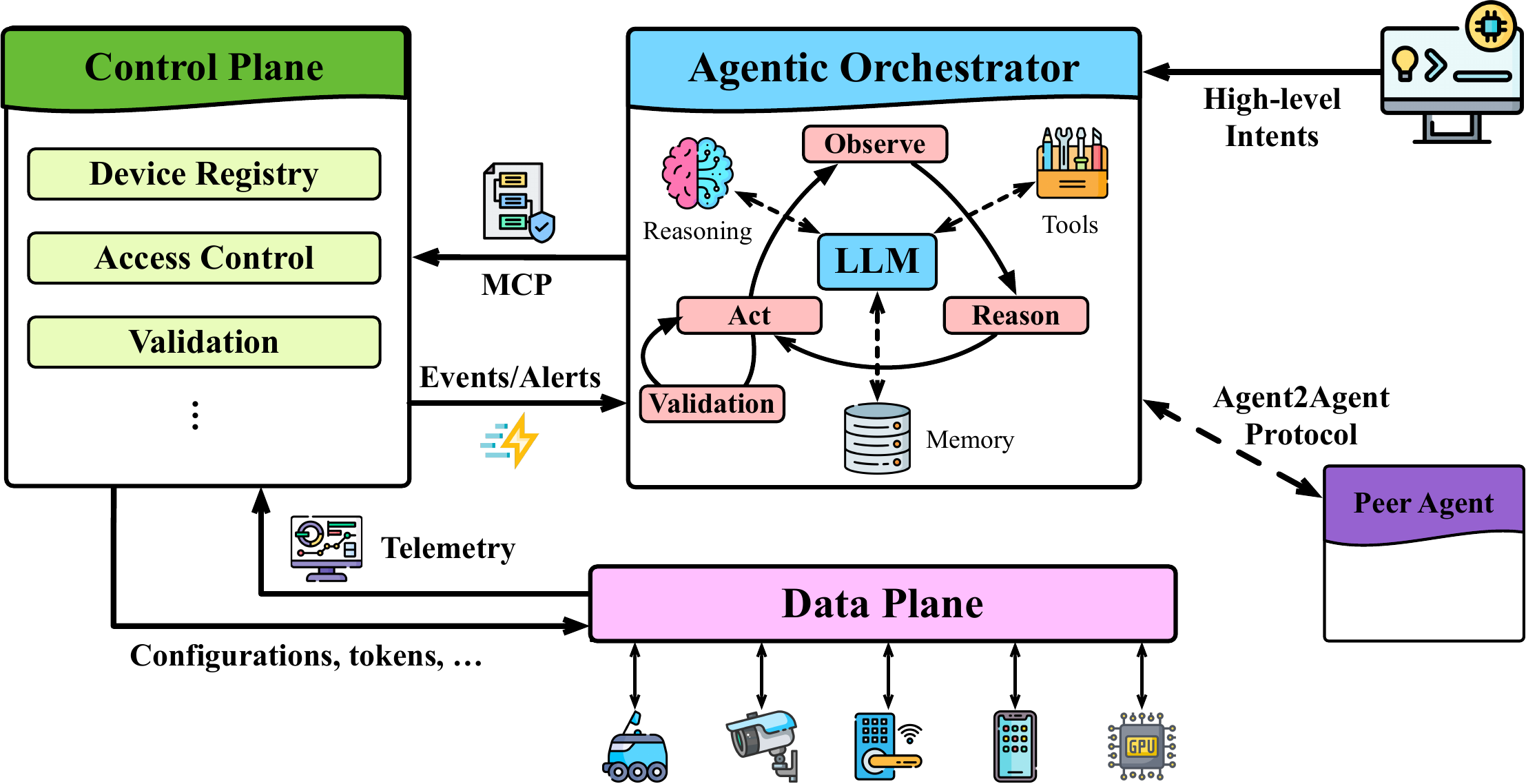}
  \caption{Reference architecture of the IoCT agentic orchestrator.}
  \label{fig:agent_arch}
\end{figure}

\cref{fig:agent_arch} illustrates the reference architecture for the IoCT
agentic orchestrator. Human administrators define high-level standing
intents (\eg \enquote{delivery robots may use lobby cameras,
but not residential devices}), while the control plane maintains the device
registry, policy state, credential and attestation state, and token issuer. The data
plane reports telemetry to the control plane and receives configurations,
policies, and capability tokens for enforcement. The agent is triggered by
events and alerts from the control plane; in response, it evaluates credential
and attestation results, administrator intent, device capabilities, resource
availability, and policy constraints to decide what access-control policies,
placements, token requests, or reconfigurations are needed.
Because these decisions affect cyber-physical resources, they must be
validated before the control plane applies them. The agent
therefore uses a multi-stage validation pipeline with syntactic and semantic
checks for generated policies, placements, and token requests, where only outputs
that pass all validation stages are applied by the control plane~\cite{Shastri2025:LLM-Driven-CollabIoT}.

The agent communicates these validated decisions to the control plane through
the Model Context Protocol (MCP)~\cite{mcp_spec_2025}, which serves as the
standardized interface between the agent and the control plane. MCP decouples
the agent's reasoning logic from
control-plane implementation, while data-plane brokers enforce the resulting
policies and capability tokens.
Finally, since each administrative domain operates its own orchestrator, the orchestrators of
the two domains must negotiate as peers when a visiting device crosses a domain
boundary, since neither has authority over the other. This inter-agent communication uses semantics \emph{inspired} by the
Agent-to-Agent (A2A) protocol~\cite{google_agent2agentprotocol2026}, where
each agent publishes an Agent Card describing its domain's capabilities
and policies, and negotiation proceeds through a structured task
lifecycle.% (propose, counter, accept, reject).

\subsection{The Collaboration Lifecycle}
\label{sec:lifecycle}
\begin{figure}[t]
  \centering
  \includegraphics[width=\linewidth]{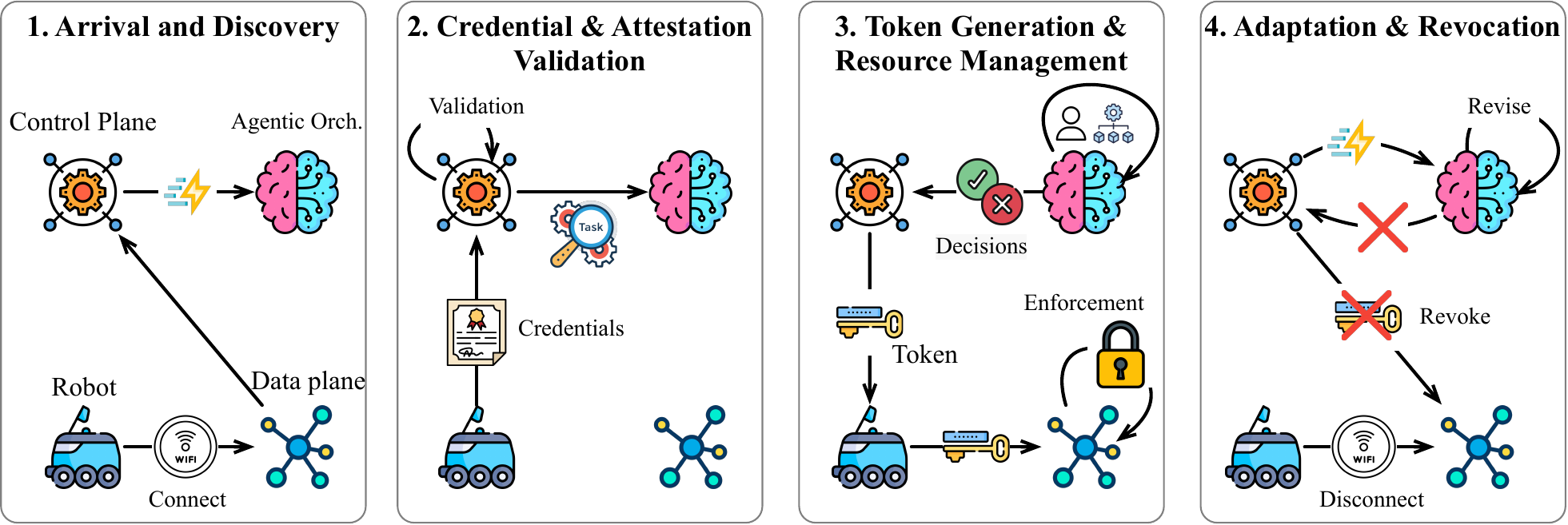}
  \caption{Collaboration lifecycle for the delivery-robot scenario.}
  \label{fig:lifecycle}
\end{figure}

\Cref{fig:lifecycle} summarizes the delivery-robot collaboration lifecycle,
from arrival to departure.
When the robot enters the
building lobby, the lifecycle begins with \emph{discovery and
capability profiling}. The building control plane then observes a newly associated
device through the deployment's discovery mechanism and raises an event to the
building's agentic orchestrator. The orchestrator queries the device registry, inspects
the robot's advertised identity, capabilities, and task needs, and retrieves
the relevant administrator intent through the MCP interface to the control
plane.

The control plane then performs the required cross-domain security checks,
including \emph{credential and attestation validation}. These checks authenticate
the robot's identity and validate claims such as its operator,
device type, declared capabilities, its firmware, and delivery purpose. The orchestrator uses
the validated identity and claims, together with the
administrator intent, current building context, available resources, and the
robot's delivery task, to derive \emph{scoped access control} and
\emph{resource placement} decisions. In this example, it may decide
that the robot can use the building's map, elevator, and lock-actuation
nanoservices, as well as on-premise compute resources needed for the delivery
task.

The control plane applies these decisions by issuing short-lived signed
capability tokens to the robot.
The robot presents these tokens when accessing resident resources, and
the data plane validates them and only permits authorized interactions. As the visit progresses,
data-plane telemetry flows back to the control plane. If the robot moves to a
different part of the building or resource availability changes, the control
plane raises an alert to the orchestrator, which evaluates the updated context
and decides whether policies, placements, or tokens should be changed. Next, \emph{runtime adaptation} keeps the data-plane enforcement state
aligned with the robot's changing physical location and the building's
current operating conditions. Finally, when the robot departs, the control plane triggers
\emph{revocation}, invalidating any visit-specific policies and tokens.

\section{Open Research Challenges}
\label{sec:challenges}

Realizing the IoCT vision poses research challenges that span the
device-layer substrate, cross-domain security, and autonomous
orchestration.

\smallskip\noindent\textbf{Safe and Verifiable Agentic Decisions.}
The agentic orchestrator derives resource-management and access-control decisions from context, device capabilities, and domain policies, while the control plane applies them. When implemented with LLM-based reasoning, however, the orchestrator may generate hallucinated or inconsistent outputs~\cite{Huang2025:LLMHallucination}, risking excessive access, incorrect placement, or unsafe actuation. Existing LLM-based policy generation and validation pipelines address parts of this problem but do not guarantee correctness for broader IoCT decisions~\cite{sonune2025lmn, vatsa2025synthesizing,cheng2025:SayWhatYouMean,jayasundara2024sok,Shastri2025:LLM-Driven-CollabIoT}. Future work must combine structured output schemas, satisfiability checks, and formal safety verification.

\smallskip\noindent\textbf{Cross-Domain Trust Negotiation.}
When a visiting device crosses a domain boundary, the two domains must establish trust and agree on resource-sharing terms without a prior relationship. Unlike traditional federated systems with long-lived agreements, IoCT interactions are transient and may involve heterogeneous credentials, attestations, and policy models. Existing trust-negotiation and agent-to-agent protocols do not cover the credential exchange, attestation validation, and policy reconciliation needed by IoCT~\cite{Winsborough2000:TrustNegotiation,google_agent2agentprotocol2026}. Open questions include resolving policy conflicts, scoping delegation, and assigning accountability when a visiting device causes harm.

\smallskip\noindent\textbf{Bidirectional Resource Sharing Under Asymmetric Trust.}
IoCT enables visiting devices not only to consume host resources but also to contribute capabilities, such as a robot GPU accelerating building inference tasks. This creates asymmetric trust, since the host may benefit from a visitor's resources without trusting it like local infrastructure. Existing isolation and cryptographic mechanisms are often hardware-specific, unavailable on many devices, or too costly for latency-sensitive workloads~\cite{Brasser2015:TyTAN,Acar2018:HESurvey}. Future systems must support varying levels of trust~\cite{RFC8485}, partition workloads to protect sensitive state, verify results from less-trusted executors, and update trust levels as evidence changes during a visit.

\smallskip\noindent\textbf{Dynamic and Heterogeneous Resource Management.}
IoCT environments are fluid, with devices arriving, departing, moving between zones, and failing without warning. The device pool is also highly heterogeneous, spanning microcontrollers, embedded processors, GPUs, and specialized accelerators that often lack standard virtualization and resource-management abstractions. The control plane must therefore re-orchestrate placement, policies, and resource allocations under latency and safety constraints as workloads and device availability change. Open challenges include managing execution state after unexpected departures and sharing sensors, actuators, and accelerators across tenants without conventional memory-protection mechanisms.

\smallskip\noindent\textbf{Incentives and Economic Models for Collaboration.}
IoCT assumes devices and environments will voluntarily contribute sensing, compute, networking, and actuation resources to transient collaborations. However, these resources incur energy, wear, privacy, and opportunity costs. Future systems require incentive and accounting mechanisms to quantify resource contributions, prioritize competing workloads, and prevent abuse or free-riding behavior. Open questions include how to value resource usage, enforce fairness, and choose among cooperative, contractual, or market-driven models.

\section{Conclusion}
\label{sec:conclusion}

% The IoT is approaching a scale and mobility regime that static,
% human-configured orchestration cannot sustain. As devices become more
% capable and mobile, they must collaborate directly with nearby devices
% across administrative trust boundaries, pooling compute, sensing, and
% actuation into dynamic clusters rather than offloading to centralized
% infrastructure. This paper has defined the Internet of Collaborating
% Things as a horizontal paradigm for this emerging reality, proposed an
% initial architecture combining a portable execution substrate with an
% LLM-driven agentic control plane, and identified open research challenges
% spanning verifiable agentic decisions, cross-domain trust negotiation,
% bidirectional resource sharing under asymmetric trust, and resource
% management over heterogeneous and dynamic device populations. 

In this position paper, we presented the Internet of Collaborating Things, a new paradigm that shifts IoT devices from isolated endpoints to autonomous participants in a continuously evolving computational collective.

\section*{Acknowledgement}\label{sec:acknowledgements}
This research was supported by National Science Foundation (NSF) grants 2213636, 2211302, 2211888, 2325956, 210549, and 2211301; U.S. Army grant W911NF-17-2-0196; Sandia National Laboratories award 2169310; and National Institutes of Health (NIH) award P41EB028242.

\bibliographystyle{IEEEtran}
\bibliography{main}

@INPROCEEDINGS{Devalal2018:LoRaComm,
  author={Devalal, Shilpa and Karthikeyan, A.},
  booktitle={2018 Second International Conference on Electronics, Communication and Aerospace Technology (ICECA)}, 
  title={{LoRa Technology - An Overview}}, 
  year={2018},
  volume={},
  number={},
  pages={284-290},
  doi={10.1109/ICECA.2018.8474715}
}

@inproceedings{arkannezhad2024ida,
  title={{IDA: Hybrid Attestation with Support for Interrupts and TOCTOU}},
  author={Arkannezhad, Fatemeh and Feng, Justin and Sehatbakhsh, Nader},
  booktitle={Network and Distributed System Security (NDSS) Symposium},
  year={2024}
}

@inproceedings{wang2022rt,
  title={{RT-TEE: Real-time System Availability for Cyber-physical Systems using ARM TrustZone}},
  author={Wang, Jinwen and Li, Ao and Li, Haoran and Lu, Chenyang and Zhang, Ning},
  booktitle={2022 IEEE Symposium on Security and Privacy (SP)},
  pages={352--369},
  year={2022},
  organization={IEEE}
}

@inproceedings{alder2021aion,
  title={Aion: Enabling open systems through strong availability guarantees for enclaves},
  author={Alder, Fritz and Van Bulck, Jo and Piessens, Frank and M{\"u}hlberg, Jan Tobias},
  booktitle={Proceedings of the 2021 ACM SIGSAC Conference on Computer and Communications Security},
  pages={1357--1372},
  year={2021}
}

@inproceedings{yuhala2024fortress,
  title={Fortress: Securing iot peripherals with trusted execution environments},
  author={Yuhala, Peterson and M{\'e}n{\'e}trey, J{\"a}mes and Felber, Pascal and Pasin, Marcelo and Schiavoni, Valerio},
  booktitle={Proceedings of the 39th ACM/SIGAPP Symposium on Applied Computing},
  pages={243--250},
  year={2024}
}

@inproceedings{menetrey2022watz,
  title={Watz: A trusted webassembly runtime environment with remote attestation for trustzone},
  author={M{\'e}n{\'e}trey, J{\"a}mes and Pasin, Marcelo and Felber, Pascal and Schiavoni, Valerio},
  booktitle={2022 IEEE 42nd International Conference on Distributed Computing Systems (ICDCS)},
  pages={1177--1189},
  year={2022},
  organization={IEEE}
}

@article{noorman2017sancus,
  title={{Sancus 2.0: A Low-Cost Security Architecture for IoT Devices}},
  author={Noorman, Job and Bulck, Jo Van and M{\"u}hlberg, Jan Tobias and Piessens, Frank and Maene, Pieter and Preneel, Bart and Verbauwhede, Ingrid and G{\"o}tzfried, Johannes and M{\"u}ller, Tilo and Freiling, Felix},
  journal={ACM Transactions on Privacy and Security (TOPS)},
  volume={20},
  number={3},
  pages={1--33},
  year={2017},
  publisher={ACM New York, NY, USA}
}

@patent{schussmann2017bluetooth,
  title        = {{Bluetooth low energy (BLE) communication between a mobile device and a vehicle}},
  number       = {US20170164192A1},
  author       = {Schussmann, Jennifer J. and Saxton, Lynn and Testa, Alessandro and Sayre, David K. and Leboeuf, Karl B.},
  assignee     = {GM Global Technology Operations LLC},
  year         = {2017},
  month        = jun,
  day          = {8},
  nationality  = {US},
  type         = {Patent Application Publication},
  url          = {https://patents.google.com/patent/US20170164192A1/en}
}

@online{aws-iot,
  author  = {{Amazon Web Services}},
  title   = {{AWS IoT}},
  url     = {https://aws.amazon.com/iot/},
  year    = {2026},
  urldate = {2026-04-20}
}

@misc{aws_wavelength_2026,
  author       = {{Amazon Web Services, Inc.}},
  title        = {{5G Edge Computing Infrastructure -- AWS Wavelength}},
  year         = {2026},
  howpublished = {\url{https://aws.amazon.com/wavelength/}},
  note         = {Accessed: 2026-04-20}
}

@misc{azure_iot_edge_2026,
  author       = {{Microsoft}},
  title        = {{Azure IoT Edge Documentation} | {Microsoft Learn}},
  year         = {2026},
  howpublished = {\url{https://learn.microsoft.com/en-us/azure/iot-edge/}},
  note         = {Accessed: 2026-04-20}
}

@misc{google_agent2agentprotocol2026,
  title        = {Agent2Agent Protocol},
  author       = {{A2A Project}},
  year         = {2026},
  howpublished = {\url{https://a2a-protocol.org/latest/}},
  note         = {Accessed: 2026-04-24}
}

@inproceedings{Shastri2025:LLM-Driven-CollabIoT,
  author    = {Shastri, Hetvi and Hanafy, Walid and Wu, Li and Irwin, David and Srivastava, Mani and Shenoy, Prashant},
  title     = {{LLM-Driven Auto Configuration for Transient IoT Device Collaboration}},
  year      = {2025},
  isbn      = {9798400722387},
  publisher = {Association for Computing Machinery},
  address   = {New York, NY, USA},
  url       = {https://doi.org/10.1145/3769102.3770619},
  doi       = {10.1145/3769102.3770619},
  booktitle = {Proceedings of the Tenth ACM/IEEE Symposium on Edge Computing},
  articleno = {7},
  numpages  = {17},
  location  = {the Hilton Arlington National Landing, Arlington, VA, USA},
  series    = {SEC '25}
}

@article{Huang2025:LLMHallucination,
  author     = {Huang, Lei and Yu, Weijiang and Ma, Weitao and Zhong, Weihong and Feng, Zhangyin and Wang, Haotian and Chen, Qianglong and Peng, Weihua and Feng, Xiaocheng and Qin, Bing and Liu, Ting},
  title      = {{A Survey on Hallucination in Large Language Models: Principles, Taxonomy, Challenges, and Open Questions}},
  year       = {2025},
  issue_date = {March 2025},
  publisher  = {Association for Computing Machinery},
  address    = {New York, NY, USA},
  volume     = {43},
  number     = {2},
  issn       = {1046-8188},
  url        = {https://doi.org/10.1145/3703155},
  doi        = {10.1145/3703155},
  journal    = {ACM Trans. Inf. Syst.},
  month      = jan,
  articleno  = {42},
  numpages   = {55}
}

@misc{IoTAnalytics2025,
  author       = {{IoT Analytics}},
  title        = {{State of IoT 2025: Number of connected IoT devices growing 14\% to 21.1 billion globally}},
  howpublished = {\url{https://iot-analytics.com/number-connected-iot-devices/}},
  year         = {2025},
  month        = {October},
  note         = {Accessed: 2026-04-20}
}

@misc{sonune2025lmn,
  title         = {{LMN: A Tool for Generating Machine Enforceable Policies from Natural Language Access Control Rules using LLMs}},
  author        = {Pratik Sonune and Ritwik Rai and Shamik Sural and Vijayalakshmi Atluri and Ashish Kundu},
  year          = {2025},
  eprint        = {2502.12460},
  archiveprefix = {arXiv},
  primaryclass  = {cs.CR},
  url           = {https://arxiv.org/abs/2502.12460}
}

@misc{vatsa2025synthesizing,
  title         = {{Synthesizing Access Control Policies using Large Language Models}},
  author        = {Adarsh Vatsa and Pratyush Patel and William Eiers},
  year          = {2025},
  eprint        = {2503.11573},
  archiveprefix = {arXiv},
  primaryclass  = {cs.SE},
  url           = {https://arxiv.org/abs/2503.11573}
}

@misc{cheng2025:SayWhatYouMean,
  title         = {{Say What You Mean: Natural Language Access Control with Large Language Models for Internet of Things}},
  author        = {Ye Cheng and Minghui Xu and Yue Zhang and Kun Li and Hao Wu and Yechao Zhang and Shaoyong Guo and Wangjie Qiu and Dongxiao Yu and Xiuzhen Cheng},
  year          = {2025},
  eprint        = {2505.23835},
  archiveprefix = {arXiv},
  primaryclass  = {cs.CL},
  url           = {https://arxiv.org/abs/2505.23835}
}

@misc{mcp_spec_2025,
  title        = {Model Context Protocol Specification},
  author       = {{Model Context Protocol}},
  year         = {2025},
  month        = nov,
  day          = {25},
  howpublished = {\url{https://modelcontextprotocol.io/specification/2025-11-25}},
  note         = {Accessed: 2026-04-24}
}

@article{jayasundara2024sok,
  title     = {SoK: Access Control Policy Generation from High-level Natural Language Requirements},
  author    = {Jayasundara, Sakuna Harinda and Gamagedara Arachchilage, Nalin Asanka and Russello, Giovanni},
  journal   = {ACM Computing Surveys},
  volume    = {57},
  number    = {4},
  pages     = {1--37},
  year      = {2024},
  publisher = {ACM New York, NY}
}

@article{Wang2024:LLMAgentSurvey,
  author     = {Wang, Lei and Ma, Chen and Feng, Xueyang and Zhang, Zeyu and Yang, Hao and Zhang, Jingsen and Chen, Zhiyuan and Tang, Jiakai and Chen, Xu and Lin, Yankai and Zhao, Wayne Xin and Wei, Zhewei and Wen, Jirong},
  doi        = {10.1007/s11704-024-40231-1},
  isbn       = {2095-2236},
  journal    = {Frontiers of Computer Science},
  number     = {6},
  pages      = {186345},
  title      = {A survey on large language model based autonomous agents},
  url        = {https://doi.org/10.1007/s11704-024-40231-1},
  volume     = {18},
  year       = {2024}
}

@misc{Apple_vison_pro,
  author = {Apple Inc.},
  title  = {Apple Vision Pro - Apple},
  year   = {2024},
  url    = {https://www.apple.com/apple-vision-pro/}
}

@article{Sumers2024:CoALA,
  title     = {{Cognitive Architectures for Language Agents}},
  author    = {Sumers, \{Theodore R.\} and Shunyu Yao and Karthik Narasimhan and Griffiths, \{Thomas L.\}},
  year      = {2024},
  volume    = {2024},
  journal   = {Transactions on Machine Learning Research},
  issn      = {2835--8856},
  publisher = {Transactions on Machine Learning Research}
}

@misc{amazon_key,
  author = {Stacey Wolfson},
  title  = {{Amazon Key In-Garage Delivery: What It Is and How It Works}},
  year   = {2023},
  url    = {https://www.aboutamazon.com/news/amazon-prime/what-is-amazon-key-in-garage-delivery},
  note   = {Accessed: 2024-11-10}
}

@misc{AWSIoTGreengrass,
  author       = {AWS},
  title        = {AWS IoT Greengrass Documentation},
  howpublished = {\url{https://docs.aws.amazon.com/greengrass/}},
  note         = {Accessed: 2023-10-23},
  year         = 2023
}

@inproceedings{yao2023:React,
  title     = {{ReAct: Synergizing Reasoning and Acting in Language Models}},
  author    = {Yao, Shunyu and Zhao, Jeffrey and Yu, Dian and Du, Nan and Shafran, Izhak and Narasimhan, Karthik and Cao, Yuan},
  booktitle = {International Conference on Learning Representations (ICLR) },
  year      = {2023},
  html      = {https://arxiv.org/abs/2210.03629}
}

@inproceedings{Shen2023:HuggingGPT,
  author    = {Shen, Yongliang and Song, Kaitao and Tan, Xu and Li, Dongsheng and Lu, Weiming and Zhuang, Yueting},
  title     = {HuggingGPT: solving AI tasks with chatgpt and its friends in hugging face},
  year      = {2023},
  publisher = {Curran Associates Inc.},
  address   = {Red Hook, NY, USA},
  booktitle = {Proceedings of the 37th International Conference on Neural Information Processing Systems},
  articleno = {1657},
  numpages  = {27},
  location  = {New Orleans, LA, USA},
  series    = {NIPS '23}
}

@inproceedings{Meng2022:DoWeNeedEdge,
  author    = {Meng, Jiayi and Kong, Z. Jonny and Hu, Y. Charlie and Choi, Mun Gi and Lal, Dhananjay},
  title     = {{Do We Need Sophisticated System Design for Edge-assisted Augmented Reality?}},
  year      = {2022},
  isbn      = {9781450392532},
  url       = {https://doi.org/10.1145/3517206.3526267},
  doi       = {10.1145/3517206.3526267},
  booktitle = {Proceedings of the 5th International Workshop on Edge Systems, Analytics and Networking},
  pages     = {7–-12},
  numpages  = {6},
  location  = {Rennes, France},
  series    = {EdgeSys '22}
}

@article{achir2022service,
  title   = {{Service discovery and selection in IoT: A survey and a taxonomy}},
  journal = {Journal of Network and Computer Applications},
  volume  = {200},
  pages   = {103331},
  year    = {2022},
  issn    = {1084-8045},
  doi     = {https://doi.org/10.1016/j.jnca.2021.103331},
  author  = {Meriem Achir and Abdelkrim Abdelli and Lynda Mokdad and Jalel Benothman}
}

@article{Shakarami2022ScenarioDrivenDA,
  title   = {Scenario-Driven Device-to-Device Access Control in Smart Home IoT},
  author  = {Mehrnoosh Shakarami and James O. Benson and Ravi S. Sandhu},
  journal = {2022 IEEE 4th International Conference on Trust, Privacy and Security in Intelligent Systems, and Applications (TPS-ISA)},
  year    = {2022},
  pages   = {217-228},
  url     = {https://api.semanticscholar.org/CorpusID:257537545}
}

@inproceedings{Wei2022:CoT,
  author    = {Wei, Jason and Wang, Xuezhi and Schuurmans, Dale and Bosma, Maarten and Ichter, Brian and Xia, Fei and Chi, Ed H. and Le, Quoc V. and Zhou, Denny},
  title     = {{Chain-of-Thought Prompting Elicits Reasoning
in Large Language Models}},
  year      = {2022},
  isbn      = {9781713871088},
  publisher = {Curran Associates Inc.},
  address   = {Red Hook, NY, USA},
  booktitle = {Proceedings of the 36th International Conference on Neural Information Processing Systems},
  articleno = {1800},
  numpages  = {14},
  location  = {New Orleans, LA, USA},
  series    = {NIPS '22}
}

@inproceedings{Liu2021:Aerogel,
  author    = {Liu, Renju and Garcia, Luis and Srivastava, Mani},
  booktitle = {{2021 IEEE/ACM Symposium on Edge Computing (SEC)}},
  title     = {{Aerogel: Lightweight Access Control Framework for WebAssembly-Based Bare-Metal IoT Devices}},
  year      = {2021},
  volume    = {},
  number    = {},
  pages     = {94-105},
  doi       = {10.1145/3453142.3491282}
}

@inproceedings{Lin2020:MCUNet,
  author    = {Lin, Ji and Chen, Wei-Ming and Lin, Yujun and Cohn, John and Gan, Chuang and Han, Song},
  title     = {{MCUNet: Tiny Deep Learning on IoT Devices}},
  year      = {2020},
  isbn      = {9781713829546},
  publisher = {Curran Associates Inc.},
  address   = {Red Hook, NY, USA},
  booktitle = {Proceedings of the 34th International Conference on Neural Information Processing Systems},
  articleno = {982},
  numpages  = {12},
  location  = {Vancouver, BC, Canada},
  series    = {NIPS '20}
}

@techreport{ocf_upnp_arch_2020,
  author      = {{Open Connectivity Foundation}},
  title       = {{UPnP} Device Architecture 2.0},
  institution = {Open Connectivity Foundation},
  year        = {2020},
  month       = {April},
  url         = {https://openconnectivity.org/upnp-specs/UPnP-arch-DeviceArchitecture-v2.0-20200417.pdf},
  note        = {Accessed: 2026-04-20}
}

@misc{IoTTrillion,
  author       = {Ian Scales},
  title        = {ARM predicts 1 trillion {IoT} devices by 2035 with new end-to-end platform},
  year         = {2020},
  month        = {April},
  publisher    = {International Telecommunication Union (ITU)},
  howpublished = {\url{https://www.itu.int/hub/2020/04/arm-predicts-1-trillion-iot-devices-by-2035-with-new-end-to-end-platform/}},
  note         = {Accessed: 2026-04-20}
}

@article{Alkhresheh2020:DACIoT,
  author   = {Alkhresheh, Ashraf and Elgazzar, Khalid and Hassanein, Hossam S.},
  journal  = {IEEE Internet of Things Journal},
  title    = {{DACIoT: Dynamic Access Control Framework for IoT Deployments}},
  year     = {2020},
  volume   = {7},
  number   = {12},
  pages    = {11401-11419},
  doi      = {10.1109/JIOT.2020.3002709}
}

@inproceedings{Sehatbakhsh2019:EMMA,
  author    = {Sehatbakhsh, Nader and Nazari, Alireza and Khan, Haider and Zajic, Alenka and Prvulovic, Milos},
  title     = {{EMMA: Hardware/Software Attestation Framework for Embedded Systems Using Electromagnetic Signals}},
  year      = {2019},
  isbn      = {9781450369381},
  publisher = {Association for Computing Machinery},
  address   = {New York, NY, USA},
  url       = {https://doi.org/10.1145/3352460.3358261},
  doi       = {10.1145/3352460.3358261},
  booktitle = {Proceedings of the 52nd Annual IEEE/ACM International Symposium on Microarchitecture},
  pages     = {983–995},
  numpages  = {13},
  location  = {Columbus, OH, USA},
  series    = {MICRO-52}
}

@article{Gupta2019:Chiplets,
  author     = {Gupta, Puneet and Iyer, Subramanian S.},
  title      = {{Goodbye, motherboard. Bare chiplets bonded to silicon will make computers smaller and more powerful: Hello, silicon-interconnect fabric}},
  year       = {2019},
  issue_date = {Oct. 2019},
  publisher  = {IEEE Press},
  volume     = {56},
  number     = {10},
  issn       = {0018-9235},
  url        = {https://doi.org/10.1109/MSPEC.2019.8847587},
  doi        = {10.1109/MSPEC.2019.8847587},
  journal    = {IEEE Spectr.},
  month      = oct,
  pages      = {28--33},
  numpages   = {6}
}

@inproceedings{Noor2019:DDFlow,
  author    = {Noor, Joseph and Tseng, Hsiao-Yun and Garcia, Luis and Srivastava, Mani},
  title     = {{DDFlow: Visualized Declarative Programming for Heterogeneous IoT Networks}},
  year      = {2019},
  isbn      = {9781450362832},
  publisher = {Association for Computing Machinery},
  address   = {New York, NY, USA},
  url       = {https://doi.org/10.1145/3302505.3310079},
  doi       = {10.1145/3302505.3310079},
  booktitle = {Proceedings of the International Conference on Internet of Things Design and Implementation},
  pages     = {172--177},
  numpages  = {6},
  location  = {Montreal, Quebec, Canada},
  series    = {IoTDI '19}
}

@inproceedings{Kotsifakou2018:HPVM,
  author    = {Kotsifakou, Maria and Srivastava, Prakalp and Sinclair, Matthew D. and Komuravelli, Rakesh and Adve, Vikram and Adve, Sarita},
  title     = {{HPVM: Heterogeneous Parallel Virtual Machine}},
  year      = {2018},
  isbn      = {9781450349826},
  publisher = {Association for Computing Machinery},
  address   = {New York, NY, USA},
  url       = {https://doi.org/10.1145/3178487.3178493},
  doi       = {10.1145/3178487.3178493},
  booktitle = {Proceedings of the 23rd ACM SIGPLAN Symposium on Principles and Practice of Parallel Programming},
  pages     = {68--80},
  numpages  = {13},
  location  = {Vienna, Austria},
  series    = {PPoPP '18}
}

@inproceedings{He2018:Rethinking,
  author    = {Weijia He and Maximilian Golla and Roshni Padhi and Jordan Ofek and Markus D{\"u}rmuth and Earlence Fernandes and Blase Ur},
  title     = {{Rethinking Access Control and Authentication for the Home Internet of Things (IoT)}},
  booktitle = {27th USENIX Security Symposium (USENIX Security 18)},
  year      = {2018},
  isbn      = {978-1-939133-04-5},
  address   = {Baltimore, MD},
  pages     = {255--272},
  month     = aug
}

@article{Satya2017:Edge,
  author  = {Satyanarayanan, Mahadev},
  journal = {Computer},
  title   = {{The Emergence of Edge Computing}},
  year    = {2017},
  volume  = {50},
  number  = {1},
  pages   = {30-39},
  doi     = {10.1109/MC.2017.9}
}

@inproceedings{Haas2017:WebAssembly,
  author    = {Haas, Andreas and Rossberg, Andreas and Schuff, Derek L. and Titzer, Ben L. and Holman, Michael and Gohman, Dan and Wagner, Luke and Zakai, Alon and Bastien, JF},
  title     = {{Bringing the Web up to Speed with WebAssembly}},
  year      = {2017},
  isbn      = {9781450349888},
  publisher = {Association for Computing Machinery},
  address   = {New York, NY, USA},
  url       = {https://doi.org/10.1145/3062341.3062363},
  doi       = {10.1145/3062341.3062363},
  booktitle = {Proceedings of the 38th ACM SIGPLAN Conference on Programming Language Design and Implementation},
  pages     = {185--200},
  numpages  = {16},
  location  = {Barcelona, Spain},
  series    = {PLDI 2017}
}

@article{Shi2016:EdgeComputingVision,
  author   = {Shi, Weisong and Cao, Jie and Zhang, Quan and Li, Youhuizi and Xu, Lanyu},
  journal  = {IEEE Internet of Things Journal},
  title    = {{Edge Computing: Vision and Challenges}},
  year     = {2016},
  volume   = {3},
  number   = {5},
  pages    = {637--646},
  doi      = {10.1109/JIOT.2016.2579198}
}

@inproceedings{ccori2016device,
  title        = {{Device discovery strategies for the IoT}},
  author       = {Ccori, Pablo Calcina and De Biase, Laisa Caroline Costa and Zuffo, Marcelo Knorich and da Silva, Fl{\'a}vio Soares Corr{\^e}a},
  booktitle    = {2016 IEEE International Symposium on Consumer Electronics (ISCE)},
  pages        = {97--98},
  year         = {2016},
  organization = {IEEE}
}

@inproceedings{Brasser2015:TyTAN,
  author    = {Brasser, Ferdinand and El Mahjoub, Brahim and Sadeghi, Ahmad-Reza and Wachsmann, Christian and Koeberl, Patrick},
  booktitle = {2015 52nd ACM/EDAC/IEEE Design Automation Conference (DAC)},
  title     = {{TyTAN: Tiny Trust Anchor for Tiny Devices}},
  year      = {2015},
  pages     = {1-6},
  doi       = {10.1145/2744769.2744922}
}

@inproceedings{Giang2015:Distributed,
  author    = {Giang, Nam Ky and Blackstock, Michael and Lea, Rodger and Leung, Victor C.M.},
  booktitle = {2015 5th International Conference on the Internet of Things (IOT)},
  title     = {{Developing IoT applications in the Fog: A Distributed Dataflow Approach}},
  year      = {2015},
  pages     = {155-162},
  doi       = {10.1109/IOT.2015.7356560}
}

@article{Dwork2014:AlgorithmicPrivacy,
  title     = {{The Algorithmic Foundations of Differential Privacy}},
  author    = {Dwork, Cynthia and Roth, Aaron},
  journal   = {Foundations and Trends in Theoretical Computer Science},
  volume    = {9},
  number    = {3-4},
  pages     = {211--487},
  year      = {2014},
  publisher = {Emerald Publishing Limited}
}

@article{Xu2014:IoT_Industry,
  author  = {Xu, Li Da and He, Wu and Li, Shancang},
  journal = {IEEE Transactions on Industrial Informatics},
  title   = {{Internet of Things in Industries: A Survey}},
  year    = {2014},
  volume  = {10},
  number  = {4},
  pages   = {2233-2243},
  doi     = {10.1109/TII.2014.2300753}
}

@article{zeroconf,
  author     = {Stirling, David and Al-Ali, Firas},
  title      = {{Zero Configuration Networking}},
  year       = {2003},
  issue_date = {June 2003},
  publisher  = {Association for Computing Machinery},
  address    = {New York, NY, USA},
  volume     = {9},
  number     = {4},
  issn       = {1528-4972},
  url        = {https://doi.org/10.1145/904080.904084},
  doi        = {10.1145/904080.904084},
  journal    = {XRDS},
  month      = jun,
  pages      = {19--23},
  numpages   = {5}
}

@inproceedings{Winsborough2000:TrustNegotiation,
  author    = {Winsborough, William H. and Seamons, Kent E. and Jones, Vicki E.},
  booktitle = {DARPA Information Survivability Conference and Exposition},
  title     = {{Automated Trust Negotiation}},
  year      = {2000},
  volume    = {2},
  doi       = {10.1109/DISCEX.2000.824965},
  url       = {https://doi.ieeecomputersociety.org/10.1109/DISCEX.2000.824965},
  publisher = {IEEE Computer Society},
  address   = {Los Alamitos, CA, USA},
  month     = Jan
}

@book{Waldo2000:Jini,
  title     = {{The Jini Specifications}},
  author    = {Waldo, Jim},
  year      = {2000},
  publisher = {Addison-Wesley Longman Publishing Co., Inc.}
}

@misc{ucla_robot_delivery,
  title  = {{UCLA Food Delivery with Starship Robots}},
  author = {{Associated Students UCLA (ASUCLA)}},
  url    = {https://www.asucla.ucla.edu/starship},
  note   = {Accessed: 2024-11-05}
}

@misc{MQTT5,
  author       = {Andrew Banks and Ed Briggs and Ken Borg and Raffaele De Benedetti},
  title        = {{MQTT Version 5.0}},
  howpublished = {\url{https://www.oasis-open.org/standard/mqtt-v5-0-cs02/}},
  year         = {2019},
  note         = {Accessed: 2026-04-29}
}

@misc{doordash2025dot,
  author       = {{DoorDash}},
  title        = {{DoorDash Unveils Dot, the Delivery Robot Powered by its Autonomous Delivery Platform to Accelerate Local Commerce}},
  year         = {2025},
  month        = sep,
  day          = {30},
  howpublished = {\url{https://about.doordash.com/en-us/news/doordash-unveils-dot}},
  note         = {Accessed: 2026-04-30}
}

@misc{amazon_prime_air,
  author       = {{Amazon}},
  title        = {{Prime Air Drone Delivery}},
  howpublished = {\url{https://www.amazon.com/Prime-Air-Drone-Delivery}},
  note         = {Accessed: 2026-04-30}
}

@article{Acar2018:HESurvey,
  author     = {Acar, Abbas and Aksu, Hidayet and Uluagac, A. Selcuk and Conti, Mauro},
  title      = {A Survey on Homomorphic Encryption Schemes: Theory and Implementation},
  year       = {2018},
  issue_date = {July 2019},
  publisher  = {Association for Computing Machinery},
  address    = {New York, NY, USA},
  volume     = {51},
  number     = {4},
  issn       = {0360-0300},
  url        = {https://doi.org/10.1145/3214303},
  doi        = {10.1145/3214303},
  journal    = {ACM Comput. Surv.},
  month      = jul,
  articleno  = {79},
  numpages   = {35}
}

@misc{RFC8485,
  author       = {Justin Richer and Leif Johansson},
  title        = {{Vectors of Trust}},
  howpublished = {RFC 8485},
  month        = oct,
  year         = {2018},
  publisher    = {RFC Editor},
  doi          = {10.17487/RFC8485},
  url          = {https://www.rfc-editor.org/rfc/rfc8485}
}

\end{document}